\documentclass{article}
\usepackage{arxiv}

\usepackage[utf8]{inputenc} 
\usepackage[T1]{fontenc}    
\usepackage{hyperref}       
\usepackage{url}            
\usepackage{booktabs}       
\usepackage{amsfonts}       
\usepackage{amsmath}
\usepackage{float}
\usepackage{nicefrac}       
\usepackage{longtable}
\usepackage{microtype}      
\usepackage{multirow}
\usepackage{graphicx}
\usepackage[font=small,labelfont=bf]{caption}
\usepackage[table,xcdraw]{xcolor}
\usepackage{graphics,amssymb,epsfig,color}
\usepackage{doi}

\title{PhenoLinker: Phenotype-Gene Link Prediction and Explanation using Heterogeneous Graph Neural Networks}

\date{\begin{center}
    \vspace{-2cm}
  {\large Jose L. Mellina Andreu\textsuperscript{1}, Luis Bernal\textsuperscript{1}, Antonio F. Skarmeta\textsuperscript{1}, Mina Ryten\textsuperscript{2,3}, Sara Álvarez\textsuperscript{4}, Alejandro Cisterna García\textsuperscript{1*}, Juan A. Botía\textsuperscript{1,3*†}}
  \vfill
   \textsuperscript{1}Departamento de Ingeniería de la Información y las Comunicaciones, Universidad de Murcia, Murcia, Spain.\\
  \textsuperscript{2}Genetics and Genomic Medicine, Great Ormond Street Institute of Child Health, University College London, London, WC1E 6BT, UK. \\
  \textsuperscript{3}Department of Neurodegenerative Diseases, UCL Institute of Neurology, London WC1E 6BT, UK \\
  \textsuperscript{4}NIMGenetics, Genómica y Medicina, SL, Madrid, Spain \\
  \textsuperscript{*} The authors contributed equally \\
  \textsuperscript{†}Corresponding author at: Departamento de Ingeniería de la Información y las Comunicaciones, Universidad de Murcia, Murcia, Spain. E-mail address: juanbot@um.es (J.A. Botía).
  \vfill
\end{center}}

\renewcommand{\headeright}{}
\renewcommand{\undertitle}{}
\renewcommand{\shorttitle}{}

\hypersetup{
pdftitle={PhenoLinker: Phenotype-Gene Link Prediction and Explanation using Heterogeneous Graph Neural Networks},
pdfsubject={q-bio.NC, q-bio.QM},
pdfauthor={David S.~Hippocampus, Elias D.~Striatum},
pdfkeywords={First keyword, Second keyword, More},
}

\begin{document}
\maketitle

\begin{abstract}
	The association of a given human phenotype to a genetic variant remains a critical challenge for biology. We present a novel system called PhenoLinker capable of associating a score to a phenotype-gene relationship by using heterogeneous information networks and a convolutional neural network-based model for graphs, which can provide an explanation for the predictions. This system can aid in the discovery of new associations and in the understanding of the consequences of human genetic variation.
\end{abstract}


\section{Introduction}

In the era of artificial intelligence (AI), deep learning has led to remarkable advancements in various fields, with examples such as AlphaFold \cite{jumperHighlyAccurateProtein2021}, GPT-4 \cite{openaiGPT4TechnicalReport2023} or the very recent AlphaMissense \cite{chengAccurateProteomewideMissense2023}. In all these advances, all input data had a tabular nature, i.e., all training examples were expressed as a regular, component-based, vector of values. However, as we confront modeling problems where some of the central data structures go beyond that static and rigid structure to become graphs, the need for specialized techniques emerges. Graph Neural Networks (GNNs) have been advancing rapidly in recent years, improving the results of classical deep learning techniques by combining the use of tabular data with graph structures, processing data that can be represented as graphs, such as social or citation networks. In the field of biology, these models are rapidly improving results in a multitude of tasks, such as the prediction of unlabeled proteins or the generation of synthetic molecules through actual molecular graph learning \cite{zhangGraphNeuralNetworks2021}, prediction of protein-phenotype associations with protein-protein associations networks \cite{liuHPODNetsDeepGraph2022}, or even for drug-target interactions \cite{zhangGraphNeuralNetwork2022}.

We are concerned with the task of predicting new associations between phenotypes and genes. Phenotypes are observable traits above the molecular level, such as anatomy and behavior \cite{deansFindingOurWay2015}. In our context, a gene-phenotype association represents a critical link in understanding genetic diseases, as it denotes that pathogenic mutations within the gene in question hold the potential to give rise to the associated phenotype \cite{botsteinDiscoveringGenotypesUnderlying2003, liGenomewideInferringGene2010}. The impact of genetics on human health is broad and challenging to measure primarily because it encompasses various types of diseases caused either entirely or partially by genetic factors such as complex disorders, chromosomal disorders or single-gene disorders \cite{brandImpactGeneticsGenomics2008, korfNatureFrequencyGenetic2019}. For instance, as indicated by certain studies that concentrate solely on rare diseases, approximately 71.9\% of which have a genetic basis, the population prevalence of rare diseases is estimated to be around 3.5–5.9\% \cite{nguengangwakapEstimatingCumulativePoint2020}. Hence, focusing exclusively on genetic rare diseases, we can estimate that the population prevalence would be in the range of approximately 2.5-4.2\%. In the realm of genetic diagnostics using whole-exome sequencing information, identifying gene-phenotype associations becomes a crucial step \cite{deisserothClinPhenExtractsPrioritizes2019, marwahaGuideDiagnosisRare2022, zemojtelEffectiveDiagnosisGenetic2014}. Here, clinicians, prompted by a patient's symptoms and phenotypes, curate a list of potential phenotypes, which, through previously documented gene-phenotype associations, ultimately lead to candidate genes. These genes then lead to causative variants or mutations. Therefore, phenotypes lead to genes and genes lead to possible candidate mutations from a comprehensive and carefully manually curated catalog. And a successful diagnostic happens if the geneticist manages to find a causal mutation given that the patient has that mutation and manifest the phenotypes associated with it, through the given genes. Therefore, the importance of comprehensive gene-phenotype associations becomes evident in the diagnostic process. When our knowledge about what we know of gene-phenotype associations in regard to the target disease is incomplete, due to unavailability or lack of reliable documentation, we fail to provide the requisite diagnostic information. As a precise diagnosis leads to a more specific and maybe personalized treatment, a successful diagnosis is paramount.  Unknown phenotype-gene associations contribute as one of the factors responsible for leaving undiagnosed approximately 60\% of the patients under suspicion of suffering a genetic disease \cite{wrightGenomicDiagnosisRare2023}. On top of that, it is also essential, when looking for the causative mutation, to keep the number of genes to inspect during the process (i.e., the so-called gene panel for the phenotypes implicated), in a specific set.  Expanding the set of genes with the aim of not excluding any possible causal mutation will result in a broader range of potential mutations. While increasing that probability marginally, the increase of work overload for the expert geneticist may be unacceptable. Consequently, achieving a balance between precision and recall proves pivotal in this undertaking \cite{raoDiagnosticYieldGenetic2023}. Therefore, there  is a need for more complete yet highly reliable gene-phenotype associations to increase the yield of genetic diagnosis without injecting an additional burden to medical  experts. 

To formally model what is relevant to describe associations between phenotypes and genes, we have to consider the Human Phenotype Ontology (HPO) \cite{kohlerHumanPhenotypeOntology2021b}. HPO is an ontology whose structure is a directed acyclic graph. It was created to describe the spectrum of all possible human phenotypic abnormalities. It is based on a standardized vocabulary to provide a unified representation of phenotypes and their relationships, mainly the is-a relation, that is used to express that phenotype a is a subtype of phenotype b if we say b is-a a. We have also considered as potentially useful other ontology-based  sources of information about molecular processes and diseases, such as the Gene Ontology (GO) \cite{ashburnerGeneOntologyTool2000, thegeneontologyconsortiumGeneOntologyKnowledgebase2023a}, the Kyoto Encyclopedia of Genes and Genomes (KEGG) \cite{kanehisaKEGGKyotoEncyclopedia2000} or the Human Disease Ontology (DO) \cite{schrimlHumanDiseaseOntology2022}. However, none of these sources include gene-phenotype associations suitable for genetic diagnosis that we can use to predict new gene-phenotype associations. Based on HPO, several gene-phenotype prediction methods have been proposed in the last few years, such as HPODNets \cite{liuHPODNetsDeepGraph2022}, CADA \cite{pengCADAPhenotypedrivenGene2021} and others \cite{notaroPredictionHumanPhenotype2017, patelGraphBasedLinkPrediction2022, shenHPO2VecLeveragingHeterogeneous2019b}.

\begin{figure}[h]
    \centering
    \includegraphics[width=\textwidth]{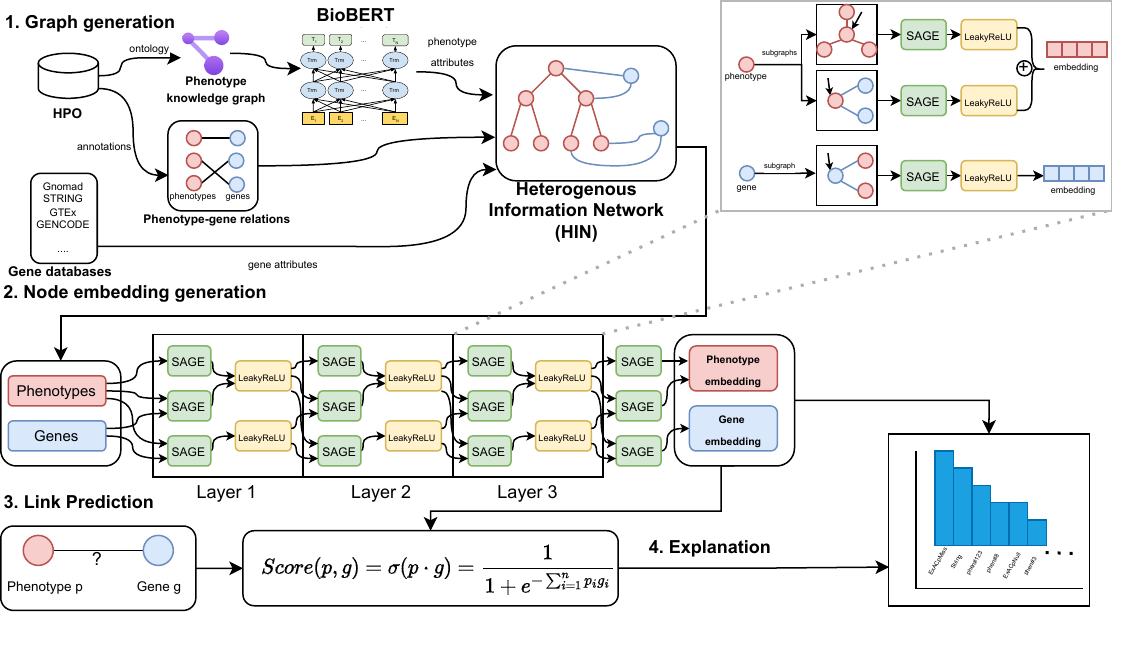}
    \vspace{-1cm}
    \caption{Overview of the overall approach for PhenoLinker. A graph is created from the HPO ontology structure which describes all phenotypes in the form of a hierarchy and includes known genes associated to phenotypes. Starting from this structure, we decorate  the corresponding gene nodes with extended features gathered from different sources reflecting different aspects of their function: genomics, gene expression and genetics. From the resulting graph, a HIN resembling that graph is created. That HIN is exploited by the HSGNN of step 2 to encode the subgraph of interest for the phenotype and the gene into vectors of real numbers (i.e., embeddings). This HSGNN was optimized to result in a SAGE based (see methods) three layer codecs for phenotype and gene. With the embeddings for both phenotype and gene, the HSGNN is then capable of predicting whether the input pair phenotype-gene should actually exist or not: the score for a given link is calculated as the sigmoid function of the dot product of gene and phenotype embeddings. Finally, the prediction can be explained on the basis of which attributes and their values were used to describe the phenotype and the gene,  with a score for each attribute reflecting its importance in the whole process. }
    \label{fig:1}
\end{figure}

Here, we have designed a totally new approach to predict gene-phenotype associations by considering Heterogeneous Information Networks (HINs), so that we have a network with nodes of different classes and with different attributes, such as phenotypes and genes, based on information from HPO. HINs are extremely interesting in this case for a number of reasons: i) Having phenotypes and genes in the same network allows treating the gene-phenotype associations seamlessly within the GNN. ii) The possibility of describing each node in the network with specific attributes leads the door open to a more detailed description of genes and phenotypes and hence, to a more comprehensible set of gene-phenotype associations and iii) the GNN training algorithm is capable of capturing how specific gene features impact the potential gene-phenotype associations. This new layer of information is beneficial for obtaining more accurate predictions, and can contribute to reach an understanding, through explainable AI, of which features drive a gene to be associated with a phenotype. Following these hypotheses, we developed a model called PhenoLinker based on heterogeneous structure graph neural networks (HSGNNs) to make the most out of this HIN (see Fig. \ref{fig:1}). Given a pair gene-phenotype used as input, the HSGNN developed in this research provides a quality score and explanations for that pair that sheds light on how the prediction was made by the neural network. The predictions of the model, scores, explanations and different visualization approaches on a certain dataset are made available to the research community in a web application we have developed via Hugging Face Spaces.

\section{Materials}

We use the December 15, 2022 version of the HPO ontology, directly available on their website, with the ontology file (hp.obo) and the file containing the gene relationships (phenotype\_to\_genes.txt). For gene attributes, information is gathered from several sources as it is done in \cite{botiaG2PUsingMachine2018}. We have gene-specific variant frequency information belonging to the Gnomad database \cite{lekAnalysisProteincodingGenetic2016}, information of gene and transcript structural complexity from GENCODE (Ensembl version 72) \cite{harrowGENCODEReferenceHuman2012} and the HEXEvent database \cite{buschHEXEventDatabaseHuman2013}, gene-specific protein-protein interactions information from the STRING database \cite{szklarczykSTRINGV10Proteinprotein2015} and tissue-specific gene expression and co-expression information from the GTEx V6 gene expression dataset. For a detailed view of the attributes see Supplementary Table \ref{tab:s1}.

A BioBERT model \cite{dekaEvidenceExtractionValidate2022} available from Hugging Face is used to create the phenotype attributes. This model was previously trained with the SNLI, MNLI, SCINLI, SCITAIL, MEDNLI, and STSB datasets, and creates 768-dimensional embeddings that we use as phenotype attributes from their textual description. The interactive online application is based on the Genomics England amber panel Intellectual disability - microarray and sequencing v5.204 \cite{IntellectualDisabilityMicroarray}.
 
For the validation of our model, we have also accessed data from the Gene2Phenotype \cite{thormannFlexibleScalableDiagnostic2019} database, using the relationships of all the panels that were not included in the used HPO version. This data was downloaded on September 20, 2023, as the data is generated on the fly with the most recent information directly from the website.

\section{Methods}

In this section, we describe the components of our model and how we applied it. We introduce the main architecture of our model, explaining the process of creating the dataset and the neural network architecture that we use to make the predictions, as well as the additional layer that implements the interpretation of the results.
 
The fundamental aim that drives our research is the discovery of new undiscovered associations between genes and phenotypes. To enable this using supervised machine learning (ML), we turn the original problem into a link prediction task on a heterogeneous graph, so that we are interested in links not belonging to the graph that our model predicts as belonging to it (i.e., the ML model predicts a high quality for that link, and subsequently, for the corresponding pair gene-phenotype). To construct the model for the link prediction task, the necessary steps are as follows: (1)  creation of the graph with phenotypes and genes emerging from HPO, the HIN, (2) creation of the HSGNN model for heterogeneous graphs capable of generating suitable embeddings for genes and phenotypes, (3) inference of scores for all possible and potentially new pairs gene-phenotype and (4) generation of interpretation of the most promising predictions. 

\subsection{HIN graph of phenotypes and genes generation}

We use HPO as the basis for our graph construction as it can be seen in the upper left part of Fig. \ref{fig:1} at Step 1. The HPO ontology stores information about phenotypes as an ontology, which can be viewed as an implicit knowledge graph with the is-a relation. We extract this knowledge graph and use it as the base graph to which we add gene nodes and edges between genes and phenotypes using the gene annotations provided also by HPO, where they incorporate in tabular form the existing phenotype-gene relationships.
 
Apart from the is-a relation between the individuals (in this case, the phenotypes), the HPO ontology also contains attributes describing the objects. Specifically, the ontology stores a textual description for each phenotype as an attribute, which we retrieve and use as the attribute of the phenotype nodes after transforming them into an embedding. For example, for the ontology term HP:0001945 (Fever), the ontology includes the description Body temperature elevated above the normal range.

To transform those descriptions into semantics preserving vectors of real values, we used a pre-trained BioBERT \cite{dekaEvidenceExtractionValidate2022} model. This natural language processing model transforms English language sentences of variable length into a 768-dimensional embedding. Since this BioBERT is pre-trained on biomedical texts, we use it without any further fine-tuning. We use the embedding as the vector of attributes of the node for the phenotype. This allows the incorporation of numerical attributes without having to generate a specific model or fine-tuning it.
 
With respect to gene nodes at the HIN, and which attributes we use to describe them we have compiled the available information and generated 64 attributes divided into 5 groups: Gnomad parameters on the genetic burden conveyed by the gene (4), gene expression across a variety of human control tissues (47 binary), parameters about the complexity of the genome locus that harbors the gene (2), the average level of interaction of the gene when translated into proteins (1) and Genomic Information (10). We designed this attribute set with the goal of indirectly describing gene function in relation with the phenotypes implied. All variables were scaled and centered before being introduced into the graph.

\section{Embedding generation for genes and phenotypes}

After we generate the graph that holds all the known gene-phenotype associations, with their corresponding attribute values, we need to appropriately learn how to encode the latent structures based on the graph that are used to encode genes and phenotypes into vectors of real numbers. In this approach, producing a score for any pair gene-phenotype is based on the cosine similarity of the corresponding embedding-based codification of both gene and phenotype. Therefore, the next step is to design codecs for genes and phenotypes that capture both the information provided through the graph structure and through the attributes of all nodes considered in their codification.

For this, we propose an HSGNN architecture based on SAGEConv \cite{hamiltonInductiveRepresentationLearning2018} layers, which is implemented in Pytorch Geometric using a new data structure called HeteroData, that allows us to create different sets of nodes with different attributes.
 
The SAGEConv operator uses a set of aggregator functions that learn to aggregate feature information from a node’s local neighborhood in the graph. Given a certain node $v$ in a graph $G= (V, E)$ we define the neighborhood as $N(v) =\{u\in V | (v,u)\in E\}$, where $V$ and $E$ are the sets of nodes and edges in the graph, respectively. The node $v$ has a certain vector of numeric attributes $x_v$ that represents the attributes of the node. The operation of the SAGEConv operator is defined as

\begin{equation*}
    x_v^{l+1}\leftarrow W_1^l\cdot x_v^l + W_2^l\cdot \text{mean}_{j\in\mathcal{N}(v)}x_j^l + \text{bias}^l,
\end{equation*}

where $l$ is the layer, $W1, W2$ are weight matrices and $\text{bias}^l$ is the bias of the linear operation. The dimensions of each matrix and the bias are determined by the input and output dimensions of the particular layer. Thus, for an input of size $n$ and an output of size $m$, the matrices $W1, W2$ belong to $M_{n\times m}$, and the bias is a vector of $m$ dimensions. 

In our architecture, we define the neighborhood in a heterogeneous graph depending on the specific subgroup of edges. Specifically, for a phenotype node, we have two subgroups of edges: those that connect the node with another phenotype and those that connect it with a gene. To handle these subgroups, we have a separate SAGEConv operator for each kind of edge. The final embedding of the phenotype nodes is generated by adding the embeddings from the layers. This means that the resulting embedding for the phenotype encodes both information from the phenotype itself and from the associated genes.
 
After each layer of SAGEConv operators, a LeakyReLU \cite{maasRectifierNonlinearitiesImprove2013} activation function is selected. The final embedding is directly generated by a final heterogeneous SAGEConv layer. The final structure can be seen in Fig. \ref{fig:1}. By empirical experimentation, it has been decided to use 3 intermediate SAGE layers followed by a final one that generates the embeddings, and 64 dimensions for the hidden channels of the model. This is shown in the results section.

\subsection{Prediction generation}

The final step involves predicting a score for a given edge in the heterogeneous graph. To achieve this, we leverage the node embeddings generated in the previous step and calculate the dot product between the embeddings. Subsequently, we use a sigmoid function to produce a probability for the given pair. 
 
Formally, given a phenotype p with a representation $(p_1, \dots ,p_n)$ and a gene $g$ with a representation $(g_1, \dots , g_n)$, the model computes the dot product and then is transformed into a probability using the sigmoid function as

\begin{equation*}
    \sigma(p\cdot g) = \frac{1}{1+e^{-\sum_{i=1}^n}p_i g_i}.
\end{equation*}

\subsection{Training and evaluating the model}

The task can be viewed as a binary classification problem for a given edge in the network. The positive class corresponds to the edges that belong to the network, and the negative class to the edges not present in the network. We use as the loss function the binary cross-entropy for each edge in the training set. Formally, for a certain vector of predictions $\hat{Y}=(\hat{y_1}, \dots , \hat{y_n})$ and the corresponding reference values Y=(y1, ... , yn), the objective function of optimization is

\begin{equation*}
    \frac{1}{n}\sum_{i=1}^n (y_i \log (\hat{y_i}) + (1-y_i)\log(1-\hat{y_i})).
\end{equation*}

This problem has some particularities derived from the fact that the potential set of negative examples (i.e., the whole set of links not present in the HPO-derived graph of phenotypes and genes), indeed has unknown positive examples that we want to detect.  In order to create a training dataset, we need positive cases (edges that are in the network) and negative cases (edges that are not). Positive cases are randomly selected from the existing links. But for the negative cases the situation is slightly more complex as in this case we focus on phenotype-gene pairs which are not connected and, from those, the absent links are used. 

As usual in supervised experiments, we need to organize the training data, i.e. the edges, into training, validation and test sets. This split is done using RandomLinkSplit from Pytorch Geometric, where an important parameter is the negative sampling ratio to include in the sets. This is the percentage of the total of potentially usable links not existing in the graph, that we may use as negative examples. Each training example in the training data is composed of a series of message passing edges and one supervision edge. Message passing edges are used by the GNN to pass the message. They are the edges through which the network passes the information in each layer. And supervision edges are used in the loss function for backpropagation. During training, message-passing edges are used to transmit the information. This means that these are the edges that will be incorporated into the SAGE layers to generate the embeddings in the model. Once the embeddings are generated, the supervision edges are used to calculate the training error and therefore, they are used to calculate the error updates that propagate backwards through the HSGNN weights. For validation, we use all the previous edges as input to the model to calculate the embeddings, and we have new validation edges that are the ones that will give us the metrics.

\section{Explanation of link predictions}

After we use the HSGNN for scoring a gene-phenotype link of interest, the model is capable of explaining the produced score in terms of the particular contribution of each node attribute, both for phenotypes and genes. This useful feature we have incorporated in our model is based on the Integrated Gradients \cite{sundararajanAxiomaticAttributionDeep2017a} algorithm, available from the Captum \cite{kokhlikyanCaptumUnifiedGeneric2020} library through Pytorch Geometric. This algorithm returns a weight associated with each attribute in reference to its importance when scoring the link between the gene and the phenotype.

For the sake of usability of our predictions, these interpretations are displayed in an interactive online application using Hugging Face Spaces, where a 3D representation of the embeddings is shown using the t-SNE \cite{maatenVisualizingDataUsing2008} method, and in addition, the 10 most important attributes are displayed with their scores for a number of genes and phenotypes. It can be accessed at \href{https://huggingface.co/spaces/Melland/PhenoLinker}{https://huggingface.co/spaces/Melland/PhenoLinker}.

\section{Results}

\subsection{Data preparation}

We used Pytorch Geometric 2.4.0 as the basis for our architecture, as it is a popular library for graph neural networks in Python. To train the neural network, we used a cross-entropy loss function, Adam optimizer, a learning ratio of 0.001, 15 epochs and a batch size of 2048. The core of our training data is the phenotypes within the HPO ontology in its December 15, 2022 version and the genes of the HPO annotations of the same version. Only the genes with no missing values in the attributes are used. Our final dataset has 16810 phenotypes, 4619 genes, 42142 relationships between phenotypes and 721540 relationships between genes and phenotypes. We define negative sampling ratio as the ratio of examples of the negative class to the positive class. Given that, we have chosen a negative sampling ratio of 4 when building the negative samples, implying 20\% of positive class examples in our training set for the link prediction task, and 30\% of supervision edges in the training set. Our estimate is that in the full model there is a negative ratio of 107. Therefore, the learning process is clearly imbalanced. To alleviate this, we subsample the negative class so the negative ratio is only 4. This choice responds to a trade-off between balancing the classes and somewhat respecting the imbalanced nature of the problem.
 
To estimate the predictive power of our model, we used the area under the precision-recall curve (AUCPR). This curve has been widely used as a preferred metric for evaluating link prediction models in large graphs \cite{garciagasullaEvaluatingLinkPrediction2015}. The link prediction task suffers from the problem of the existence of mislabeled edges (false negatives and false positives in the classification task), since a relationship could be undiscovered but should exist in the graph, or the relationship could be wrongly added. The use of the AUCPR metric partially alleviates the problem for the false negatives and is therefore preferred in these tasks. The Average Precision (AP) formula is used to calculate this area. It summarizes the curve as the weighted mean of precisions achieved at each threshold, with the increase of recall from the previous threshold used as the weight.

\begin{figure}[]
    \centering
    \includegraphics[width=\textwidth]{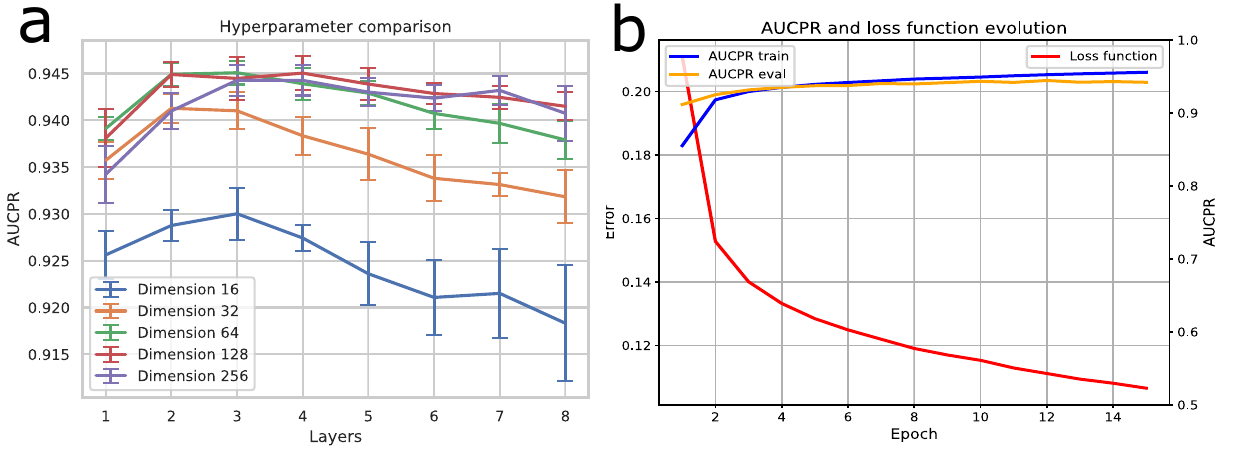}
    \caption{a) Comparative plot of model hyperparameters. The mean AUCPR of 10 random repetitions is calculated and the mean and standard deviation are taken for each combination. We have chosen 3 layers and an embedding size of 64 dimensions. b) Plot of the evolution of the AUCPR in the training and validation sets during 15 training epochs. The evolution of the loss function is also shown. Training set of 80\% and validation set of 20\% using the hyperparameters chosen from a)}
    \label{fig:2}
\end{figure}

We designed a hyperparameter grid of possible values with the number of SAGE layers that generates the embeddings of the nodes between 1 and 8, as well as with the possible dimensions of the embeddings and hidden channels of the network between 16 and 256 (see Fig.  \ref{fig:2}, panel a). The training process showed that the model did not suffer from overfitting to the training data (Fig. \ref{fig:2}, panel b). The setting was repeated 10 times with random sampling, and the mean and standard deviation are shown. Following the results, 3 layers and 64 dimensions for the embeddings were chosen, so that we have selected the simplest model without losing precision.

\subsection{Prediction performance goes beyond the state of the art.}

To compare the results of our model, we used the results obtained by HPODNets \cite{liuHPODNetsDeepGraph2022} in their paper. To the best of our knowledge, HPODNets is the deep learning-based approach that is used in a similar task to ours, and seems to be the most powerful. In the corresponding paper, its performance is compared with that of other prediction methods. This comparison is restricted to only the Phenotypic Abnormality (PA) branch terms of HPO whose phenotypes were additionally filtered leaving out those with less than 10 associated proteins through UniProt (proteins that were annotated in Swiss-Prot). With this set, the model was trained with the complete dataset, with all possible negative class edges (which implies a negative ratio of 26, that is six times greater than the one our model is trained on, i.e. 4).

Two metrics were chosen to compare our results with HPODNets: AUCPR and F1 score. We use AUCPR because it is the metric we employ thoroughly in this paper. We also use F1 because it is one of the other metrics they use. Each was reported in two ways: (i) Macro-averaged (M): calculate metrics for each term (for each gene), and calculate the mean; (ii) Micro-averaged (m): vectorize the predicted score for each pair and calculate metrics based on the resulting vector.  The dataset used has been emulated as closely as possible. We have used the December 15, 2022 version of HPO and the current UniProt and Swiss-Prot proteins to filter out the phenotypes, so we have obtained a dataframe with 4574 phenotypes and 4619 genes. On this dataset we have used the model previously trained on the global dataset, with 3 layers and 64-dimensional embeddings, using all the attributes for the nodes of the graph. 

The comparison is found in Table \ref{tab:1}, where it can be seen that our model obtains the best results by a wide margin in M-AUCPR with 50.37\%, m-AUCPR with 53.85\% and M-F1 with 64.33\%, but do not obtain the best result in m-F1 with 33.08\% whilst HPODNets obtains a 36.15\%. Since this data is really unbalanced, with a much higher proportion of negative edges, the m-F1 of our model gets much worse, as it has been designed to detect missing edges in the network and it detects many of these edges as positive. The F1 metric can be misleading for these unbalanced sets as it does not account for true negatives and false positives are not necessarily errors in this context.

\begin{table}[]
    \centering

\begin{tabular}{@{}
>{\columncolor[HTML]{FFFFFF}}c 
>{\columncolor[HTML]{FFFFFF}}c 
>{\columncolor[HTML]{FFFFFF}}c 
>{\columncolor[HTML]{FFFFFF}}c 
>{\columncolor[HTML]{FFFFFF}}c @{}}
\toprule
\textbf{Method}                           & \textbf{M-AUCPR} & \textbf{m-AUCPR} & \textbf{M-F1}    & \textbf{m-F1}    \\ \midrule
{\color[HTML]{2A2A2A} \textbf{deepNF}}    & 0.28833          & 0.24189          & 0.36646          & 0.30281          \\
{\color[HTML]{2A2A2A} \textbf{DeepMNE}}   & 0.31354          & 0.26899          & 0.39181          & 0.32425          \\
{\color[HTML]{2A2A2A} \textbf{BIONIC}}    & 0.30441          & 0.26587          & 0.38448          & 0.31973          \\
{\color[HTML]{2A2A2A} \textbf{LP}}        & 0.29329          & 0.25042          & 0.37203          & 0.30403          \\
{\color[HTML]{2A2A2A} \textbf{RANKS}}     & 0.27792          & 0.23105          & 0.36329          & 0.29188          \\
{\color[HTML]{2A2A2A} \textbf{Mashup}}    & 0.33066          & 0.29211          & 0.40789          & 0.34948          \\
{\color[HTML]{2A2A2A} \textbf{GeneMANIA}} & 0.33314          & 0.23328          & 0.40995          & 0.29647          \\
{\color[HTML]{2A2A2A} \textbf{HPODNets}}  & 0.35754          & 0.31308          & 0.43397          & \textbf{0.36152} \\
{\color[HTML]{2A2A2A} \textbf{PhenoLinker}}     & \textbf{0.50379} & \textbf{0.53855} & \textbf{0.64326} & 0.33080          \\ \bottomrule
\end{tabular}
    \caption{Performance comparison. The results of HSGNN are summarized over 5 trials. Results of previous methods are extracted from \cite{liuHPODNetsDeepGraph2022} using a weighted mean. The boldface items in the table represent the best performance. }
    \label{tab:1}
\end{table}

\subsection{Temporal validation performance is above state of the art approaches.}

We wanted to assess whether our model is able to predict absent links that should eventually be discovered in the future. Therefore we performed what we call a temporal validation. For this, we adopted a strategy where the training set is a version of HPO (December 15, 2022), and as links discovered a posteriori, we define as test sets the new and deleted links that we find in posterior versions of HPO. In this way, we can produce rough estimates of whether the model is able to predict these new additions and false relations (detected and removed by the HPO maintainers).

There are 7 versions of HPO posterior to the one we have used to train the model, although we discarded one of them (April 5, 2023) because it presents abnormally large differences with respect to the other versions (it discards 10\% of the total relationships), and it seems to be wrongly released as this change was fixed again in the following versions. In all the other cases, only the phenotypes and genes belonging to the training set are considered in the validation set. Accordingly, the links used to evaluate the model with, are those that incorporate both the new relations (positive class) and those eliminated (negative class) for those phenotypes and genes only.

The results obtained are shown in Table \ref{tab:2}. It can be seen that the version that gives the best results is the 2023-06-17 version. As time passes and the versions are later, the differences between the versions increase, and as the differences increase so does the performance of the model, because our model is trained on older data. Finally, the model behaves worse with the deleted data since the model is training with them as positively labeled edges, so this could be considered as erroneous information. We can observe an almost perfect positive correlation between the values of the AUCPR column and those for the value of the ratio of added/removed links (Pearson correlation 0.9839, P= 0.0003854).

We can compare the results with HPODNets because they also did a temporal validation too. They used as the base for training data the release of February 12, 2019. Their predictions were tested with the data release on October 12, 2020, using the m-AUCPR and M-AUCPR metrics. These metrics for the HPODNets method using the weighted mean according to term groups (they are grouped by the number of associated proteins) is 11.5\% for M-AUCPR and 8.3\% for m-AUCPR. Thus, PhenoLinker shows better results than those offered by HPODNets (see Table \ref{tab:2}).

\begin{table}[]
\centering
\resizebox{\textwidth}{!}{%
\begin{tabular}{@{}
>{\columncolor[HTML]{FFFFFF}}c 
>{\columncolor[HTML]{FFFFFF}}c 
>{\columncolor[HTML]{FFFFFF}}c 
>{\columncolor[HTML]{FFFFFF}}c 
>{\columncolor[HTML]{FFFFFF}}c 
>{\columncolor[HTML]{FFFFFF}}c 
>{\columncolor[HTML]{FFFFFF}}c 
>{\columncolor[HTML]{FFFFFF}}c @{}}
\toprule
\textbf{YY-mm-dd} & \textbf{AUCPR} & \textbf{recall} & \textbf{precision} & \textbf{F1} & \textbf{New relations} & \textbf{Removed relations} & \textbf{Ratio added/removed} \\ \midrule
2023-01-27        & 0.5906         & 0.6987          & 0.6231             & 0.6587      & 2310                   & 1039                       & 2,22                         \\
2023-06-06        & 0.7947         & 0.3872          & 0.7731             & 0.5160      & 42656                  & 5409                       & 7,88                         \\
2023-06-17        & 0.8027         & 0.3871          & 0.7806             & 0.5175      & 43059                  & 5179                       & 8,31                         \\
2023-07-21        & 0.7611         & 0.4365          & 0.7504             & 0.5520      & 62984                  & 9880                       & 6,37                         \\
2023-09-01        & 0.7449         & 0.4254          & 0.7266             & 0.5366      & 58630                  & 10136                      & 5,78                         \\
2023-10-09        & 0.7049         & 0.4289          & 0.6872             & 0.5281      & 60631                  & 12705                      & 4,77                         \\ \bottomrule
\end{tabular}%
}
\caption{Temporal validation results. AUCPR, recall, precision and F1 are shown in the sets formed with the new HPO versions. It is shown how many relationships are added and removed with respect to the training version (December 15, 2022), and the ratio between them.}
\label{tab:2}
\end{table}

\subsection{Using phenotype and gene-level attributes is effective}

\begin{figure}[]
    \centering
    \includegraphics[width=0.8\textwidth]{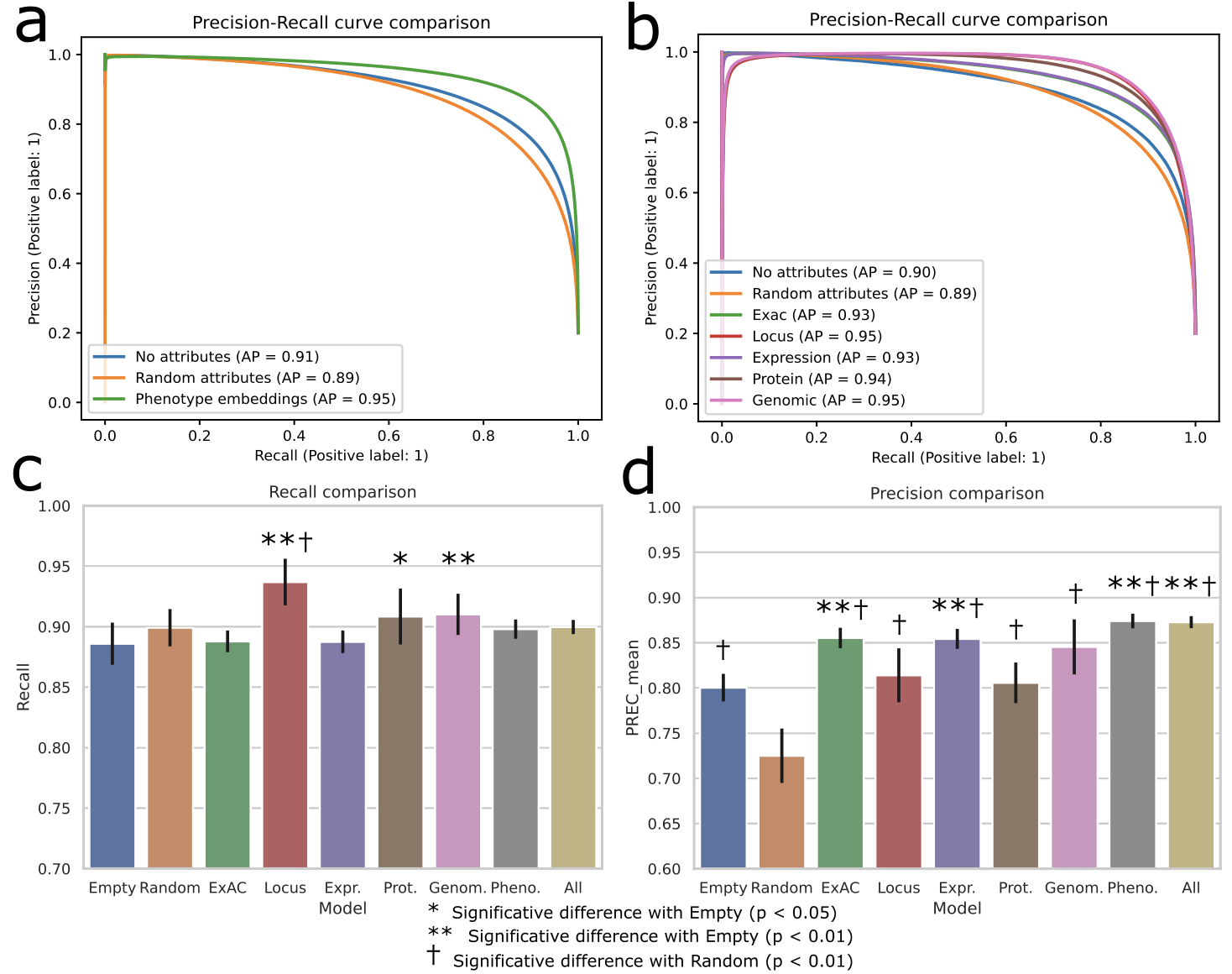}
    \caption{Comparison of models with different groups of attributes. The experiment was developed from 10 randomized repetitions with a training set of 80\% and test set of 20\%. a) We compare the precision-recall plot of the random model, a model with randomized attributes and the model where we add the embeddings of the phenotypes as attributes. AP is the Average Precision formula for the area under the Precision-Recall curve. c) and d) Graphs showing precision and recall for the different groups of attributes. The mean over the 10 replicates is plotted, and the standard deviation is shown as an error bar. The distributions were compared using Welch's t-test where we compared the means of the distributions, and symbols representing the p-value of significance are shown.}
    \label{fig:3}
\end{figure}

One of the innovations introduced by this paper is the use of attributes in the nodes of the GNN to provide a more informative description of both phenotypes and genes. To assess whether they contribute to creating better models, we faced the following questions: Is it any better to use attributes at all? Are attribute values really meaningful? What is the particular contribution of each attribute as measured separately? To try to shed some light on the answers to these questions, respectively, we estimated the precision and recall of a model with no attributes, a model with attributes whose values were assigned randomly (i.e., to simulate totally uninformative attributes) and models with a single attribute so we test how productive is each attribute alone. We followed this approach separately for the attributes at phenotypes (see Fig. \ref{fig:3}, panel a) and genes (Fig. \ref{fig:3}, panel b). 
 
Therefore, with respect to attributes at phenotypes, to assess whether the inclusion of the phenotype attributes coded into vectors of R768 through a BioBERT model is really effective in PhenoLinker, we compared it with a model with no attributes and another one with attributes with randomly assigned values, as we can see in Fig. \ref{fig:3}., panel a). We see that the AUCPR is significantly higher when we include the attributes and their values are not random. We also see that it is better not to use attributes than to use noisy ones, as we can observe with the loss of precision using random attributes. 

With respect to attributes at genes, we again compare the predictions for each group of gene attributes with empty and random models in Fig. \ref{fig:3}, panel b). All of the groups of attributes show better results than the model with no attributes, which again is better than one using random values in the attributes. The AUCPR for the random model is 89\%, for the empty model is 90\% and all models with gene attributes get an AUCPR in the interval [93-95]\%. Interestingly, the differences with respect to recall are not significant for most groups (see Fig. \ref{fig:3}, panel c), except for Locus, Protein and Genomic, but nevertheless, the precision is significantly higher for all groups except Locus and Protein (p-value < 0.01 in Welch’s t-test) with respect to empty and random models (Fig. \ref{fig:3}, panel d). This suggests that the model trained with the gene attributes is more reliable, as it obtains fewer false positives with similar recall. Thus, for PhenoLinker we incorporate both phenotype and all gene attributes, as we see that this improves the results.

\subsection{PhenoLinker predicts valid phenotype-gene associations in an alternative dataset}

To add more evidence of the model's capabilities beyond using HPO as the source for validation, we used an additional source of information to further assess the quality of PhenoLinker predictions: the Gene2Phenotype dataset.

Gene2Phenotype (G2P) \cite{thormannFlexibleScalableDiagnostic2019} is an online system containing evidence-based datasets linking genes, disease and phenotypes. In these datasets, there are relationships that have not been included in HPO. Therefore we could use them to test the efficiency of our model in predicting these new associations and to see if the model is able to infer them just from the HPO information. G2P is divided into six disease panels: Cancer, Cardiac, Developmental Disorder (DD), Eye, Skeletal and Skin. We ruled out the Cancer panel of our analyses as the size of its associations is rather small. Out of the 5829 associations in Gene2Phenotype not included in HPO, 3366 are scored by PhenoLinker with confidence to be included into HPO (i.e., their PhenoLinker score is greater than 0.5).

To get a proper assessment of whether PhenoLinker behaves better than random chance on the G2P associations, we determined whether PhenoLinker scores generated on G2P associations are higher than random ones. For such purpose, we defined two levels of randomness:  (1)  full random phenotype-gene pairs when randomly sampling from the whole HPO set of genes and phenotypes, not necessarily included into G2P and (2) random phenotype-gene pairs created by randomly sampling phenotypes and genes included in the corresponding G2P panel.  Supplementary Table \ref{tab:s3} illustrates the fold enrichment analysis we obtain in the comparison. For the full random case, the best fold is obtained for the skeletal panel. PhenoLinker score is 19 times higher when working on genuine associations than on fully random ones. The lowest is found in the DD panel, with a fold of 14. When comparing genuine associations with those randomly created using phenotypes and genes used by G2P, the highest fold is of 4.9 for the DD panel, while the lowest is found in the skeletal panel with a value of 2.5.

After a careful inspection of the PhenoLinker predictions with high confidence (>= 0.9 score) on the DD panel links, we found really remarkable hits. For example, our model found that PTEN is associated with delayed speech and language development (HP:0000750). This relationship is supported by several publications \cite{cummingsBehaviouralPsychologicalFeatures2022, steelePsychiatricCharacteristicsIndividuals2021}. In fact, this relationship has been added in a later version of HPO too. As another interesting example, our model also found that RPS23 is associated with intellectual disability (HP:0001249). This is also supported by the study presented in \cite{paoliniRibosomopathyRevealsDecoding2017}. 

\subsection{PhenoLinker predictions improve the efficiency of a genetic diagnostic application}

Thanks to the application of PhenoLinker, we have been able to improve the results of a genetic diagnostic software that prioritizes variants to assist geneticists in finding the causal variants within the exome data of clinically diagnosed patients pending the genetic diagnosis. The software in its basic form assigns scores to genetic variants based on the probability that they are causing the phenotypes identified in the patient. By adding our predictions to the system, specifically those predictions with a score greater than 0.95 and reanalyzing real cases with pending genetic definitive diagnosis we have managed to identify a number of additional variants previously validated by clinicians and genetic experts. In Table \ref{tab:3} we show these 11 gene-phenotype associations and the literature that supported these relationships.

\begin{table}[]
\centering
\resizebox{\textwidth}{!}{%
\begin{tabular}{@{}
>{\columncolor[HTML]{FFFFFF}}l 
>{\columncolor[HTML]{FFFFFF}}l 
>{\columncolor[HTML]{FFFFFF}}l @{}}
\toprule
\multicolumn{1}{c}{\cellcolor[HTML]{FFFFFF}\textbf{Gene}} &
  \multicolumn{1}{c}{\cellcolor[HTML]{FFFFFF}\textbf{New phenotypes}} &
  \multicolumn{1}{c}{\cellcolor[HTML]{FFFFFF}\textbf{Evidence}} \\ \midrule
TUBGCP2 &
  Neurodevelopmental delay (HP:0012758), Global developmental delay (HP:0001263) &
  {\cite{gungorAutosomalRecessiveVariants2021, mitaniBiallelicPathogenicVariants2019}} \\
POF1B &
  Hypogonadism (HP:0000135), Hypothyroidism (HP:0000821) &
  {\cite{lacombeDisruptionPOF1BBinding2006}} \\
EXOC7 &
  Hypotonia (HP:0001252), Delayed speech and language development (HP:0000750), Language impairment (HP:0002463) &
  {\cite{coulterRegulationHumanCerebral2020}} \\
KDM3B &
  Global developmental delay (HP:0001263) &
  {\cite{dietsNovoInheritedPathogenic2019, tabakuNovelNovoPathogenic2022}} \\
TRAPPC10 &
  Neurodevelopmental delay (HP:0012758), Global developmental delay (HP:0001263), Delayed speech and language development (HP:0000750) &
  {\cite{rawlinsBiallelicVariantsTRAPPC102022, sacherTRAPPopathiesEmergingSet2019}} \\
NRXN1 &
  Neurodevelopmental delay (HP:0012758), Global developmental delay (HP:0001263) &
  {\cite{chingDeletionsNRXN1Neurexin12010, sciaccaNRXN1DeletionTwo2022}} \\
MRPS28 &
  Neurodevelopmental delay (HP:0012758), Global developmental delay (HP:0001263) &
  {\cite{pulmanMutationsMRPS28Gene2019}} \\
UCP2 &
  Obesity (HP:0001513) &
  {\cite{liUCP2GenePolymorphisms2019}} \\
OPA1 &
  Microcephaly (HP:0000252) &
  {\cite{nascaNotOnlyDominant2017}} \\
ARHGEF10 &
  Peripheral neuropathy (HP:0009830) &
  { \cite{booraAssociationCharcotMarieToothDisease2015}} \\
ZNF335 &
  Aplasia/Hypoplasia of the corpus callosum (HP:0007370) &
  {\cite{stouffsExpandingClinicalSpectrum2018}} \\ \bottomrule
\end{tabular}%
}
\caption{New gene-phenotype associations were found using the PhenoLinker model. It shows for each gene the new associated phenotypes and their descriptions, with the literature supporting the relation.}
\label{tab:3}
\end{table}

\subsection{Results and explanations are available online}

\begin{figure}[h]
    \centering
    \includegraphics[width=\textwidth]{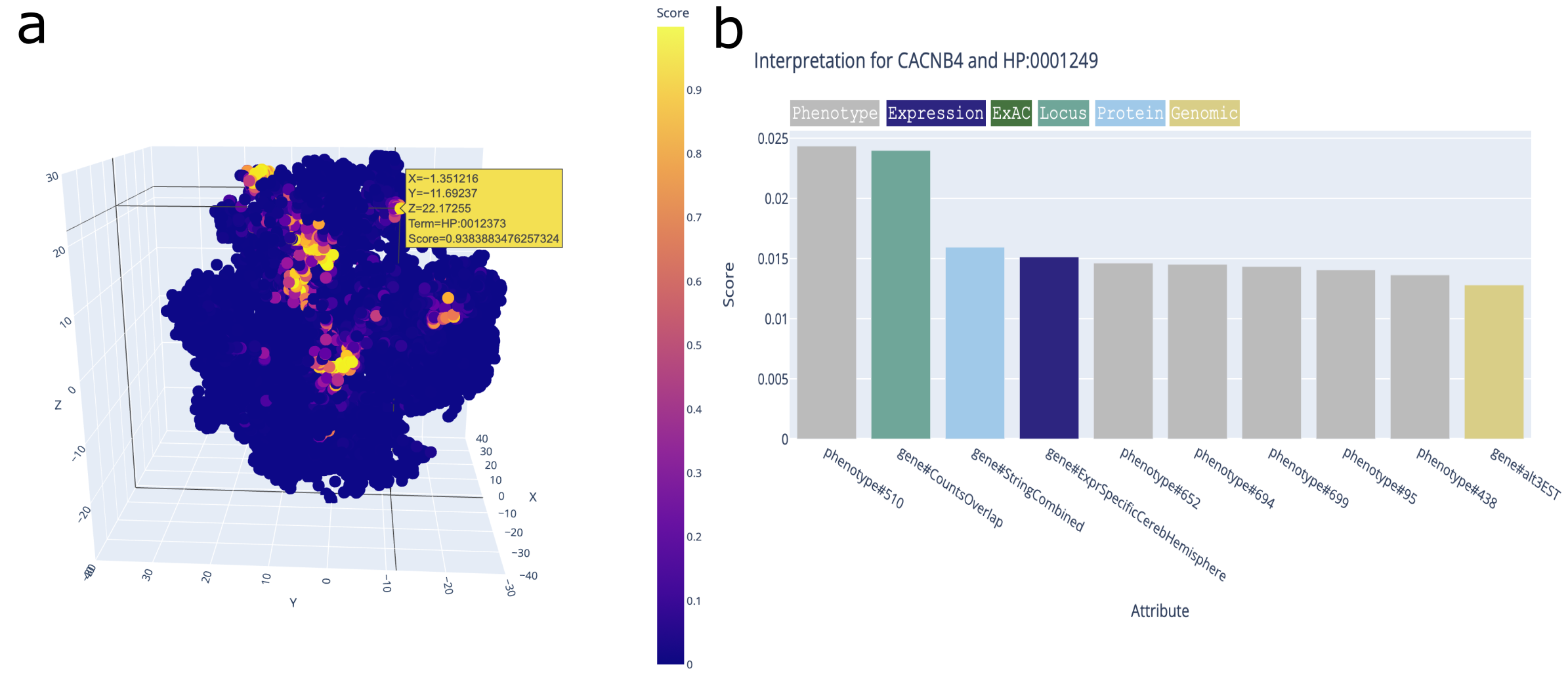}
    \caption{a) 3D representation by TSNE of the embeddings of the phenotypes, and by a heat map of the scores of these phenotypes in relation to the CACNB4 gene. It is available in the shiny application, where you can interact with the plot and select the phenotypes to see their characteristics and the specific score. b) Example of interpretation obtained for CACNB4 and phenotype HP:0001249 (Intellectual Disability) in the shiny application. The 10 most important attributes and their weights are shown in a bar chart.}
    \label{fig:4}
\end{figure}

An interactive online application for PhenoLinker is available to consult the results of a specific gene set and their corresponding linked phenotypes in Hugging Face Spaces, in this link. The application is focused on the genes from the intellectual disability Genomics England amber panel \cite{IntellectualDisabilityMicroarray}, and phenotypes have been selected such that the gene-phenotype association is not found in HPO and the score produced by the model is greater than 0.9. We selected this gene set as an interesting collection of potential genes associated with intellectual disability. PhenoLinker predicts that out of the 380 genes in the panel, 143 are linked to at least one HPO 
phenotype term associated with intellectual disability (Sup. Tables \ref{tab:s2} and 4).  One of those 143 genes is CACNB4. To illustrate the interpretability provided by our model, the Web application presents a detailed examination of the gene-phenotype relationship involving CACNB4 and the term Intellectual Disability (HP:0001249). PhenoLinker assigned a high score of 0.93 to this association, and the Web application presents the results using a bar chart that elucidates the interpretation of the findings based on the attributes of the nodes with most relevant contribution to the phenotype-gene link (see Fig \ref{fig:4}). Notably, this gene exhibits significant expression levels in the cerebellum, as supported by our data and GTEx analysis (see Sup. Fig. 1). This expression level emerged as a crucial feature in scoring this association. Furthermore, our investigation is substantiated by relevant literature discussing variants in CACNB4, intellectual disability, and cerebellar atrophy \cite{costedebagneauxHomozygousMissenseVariant2020}, and by the inclusion of this relationship in a later version of HPO (April 5, 2023).

To complement link interpretation based on contributions from each node attribute, the Web application includes an interactive 3-D representation of the embeddings for gene and phenotype, collapsed into three dimensions using t-SNE. In this way, users can visually explore the phenotype associations of a selected gene or the gene associations of a selected phenotype via a color map that highlights the relevant scores. See Fig. \ref{fig:4} for a summary of what is shown at the Web app for CACNB4. 

\section{Discussion}

In this article, we have introduced a heterogeneous structure graph neural network (HSGNN) designed for predicting gene-phenotype associations using, as training data, HPO and gene descriptions based on expression, genomics and protein interaction information, that is called PhenoLinker. The key characteristics of PhenoLinker can be summarized as follows: (1) Graph Neural Networks (GNNs): We harness GNNs to model relationships and patterns within the gene-phenotype graph. GNNs excel in capturing neighborhood information and propagating it efficiently. (2) Heterogeneous Information Networks (HINs): This implies that our network consists of nodes belonging to various classes with distinct attributes. This versatility allows us to simultaneously consider different types of connections between genes and phenotypes. (3) Utilization of specific gene attributes: This encompasses genomic and transcriptomic data sourced from a variety of origins. It enriches our model with comprehensive and diverse information for each gene. (4) Incorporating phenotype attributes: We achieve this by employing a pre-trained BioBERT to transform the detailed phenotype descriptions provided by HPO into a numerical vector of attributes using the generated embeddings.

Comprehensive experiments, which encompassed repeated random sub-sampling validation (Fig. \ref{fig:2} and Fig. \ref{fig:3}), validation similar to others (Table \ref{tab:1}), and temporal validation (Table \ref{tab:2}), lead us to get better predictions than state of the art methods. However, when we evaluated PhenoLinker´s gene-phenotype predictions like others did, to allow a fair method comparison, there are three important differences that should be considered: (1) We utilized genes instead of proteins, aligning with the HPO's representation of gene-phenotype associations. In contrast, others like HPODNets \cite{liuHPODNetsDeepGraph2022} employed proteins and mapped genes to proteins using the UniProt ID mapping tool, subsequently assigning identical associations to all proteins derived from a single gene. This approach is necessary for them due to their use of protein networks, a distinction from our methodology. (2) The HPO Version: We utilized the December 15, 2022 edition of HPO, whereas HPODNets employed the March 27, 2020, version of HPO. As of now, the March 27, 2020, version is no longer accessible. (3) Imbalanced Training: We incorporated significantly fewer false gene-phenotype associations for training our model compared to others (4 vs 26 negative sampling per positive). This discrepancy could have a detrimental effect on our evaluation metrics because, when we evaluated our model in a similar manner to others, we had to assess a considerably larger number of false associations than expected or that had been previously trained. Despite this, our results outperformed those of the previous methods.

When we assessed our temporal validation performance, we achieved superior results compared to previous methods (73.3\% for AUCPR in our model vs 11.5\% for M-AUCPR and 8.3\% for m-AUCPR in HPODNets using the averages of the results in both cases). However, it is essential to consider that different HPO versions were utilized due to the varying dates. We employed 7 versions of HPO that were released subsequent to the one utilized for training the model (from 2023-01-27 to 2023-10-09), and we assessed the performance of the model with each of these versions. This process enabled us to gather insights into the model's future performance and the connection between the inclusion and removal of gene-phenotype associations and its performance. Aside from the initial subsequent HPO version (2023-01-27), we distinctly observe a decrease in AUCPR over time (0.79 to 0.70). In terms of the impact of including and removing gene-phenotype associations, it is evident that as the negative ratio increases, the AUCPR decreases (Table \ref{tab:2}). For example, in the HPO version (2023-01-27), we observe an AUCPR of 0.59 with a negative ratio of 0.45, whereas in the HPO version (2023-06-17), we see an AUCPR of 0.80 with a reduced negative ratio of 0.12. Hence, we can deduce that the model performs better when considering new gene-phenotype associations as opposed to evaluating previously bad-included, unfavorable gene-phenotype associations in HPO. This phenomenon may be attributed to the characteristics of the training process.

The utility of PhenoLinker was further demonstrated by using valid, and reliable, phenotype-gene associations found at Gene2Phenotype but not available in HPO yet (Fig. 4). Out of 5829 associations, 3366 should be considered for inclusion in HPO following PhenoLinker. We also incorporated our predictions as additional annotations of variants in cases pending of a genetic disease diagnosis previously validated by clinicians (Table 3). By incorporating gene-phenotype associations identified by our model into a production software tool that leverages gene-phenotype information, we were able to identify an additional 11 causative variants not discovered before and confirmed by experts. Hence, this suggests that PhenoLinker may be useful for variant-ranking tools like PhenoVar \cite{trakadisPhenoVarPhenotypedrivenApproach2014}, Phrank \cite{jagadeeshPhrankMeasuresPhenotype2019}, Starvar \cite{kafkasStarvarSymptombasedTool2023},  Xrare \cite{liXrareMachineLearning2019}, DeepPVP \cite{boudelliouaDeepPVPPhenotypebasedPrioritization2019a} and commercial solutions (e.g., Emedgene, or Moon platforms).

An advantage of our model is its capacity to streamline the analysis of genes that have not yet been integrated into HPO, thanks to the inclusion of gene attributes into the model. HPO currently encompasses 4619 genes, whereas there are approximately 20,000 protein-coding genes in existence \cite{salzbergOpenQuestionsHow2018}. It is highly probable that some of the unused genes will manifest unknown links with phenotypes in HPO. At the very least, we believe that it is worth exploring this possibility in the future and assessing the accuracy of the highest quality predictions of special interest because of the phenotypes involved, through wet lab experiments, population analysis and community research.

Another novel aspect introduced in this paper was the utilization of a pre-trained BioBERT model to incorporate phenotype descriptions as attributes. We showcased that using these text descriptions coded as embeddings enhances the model's performance. But using embeddings poses new challenges. On the one hand, the BioBERT is not perfectly tailored or fine-tuned for this specific problem. We will explore fine-tuning to improve the embeddings and hopefully the PhenoLinker predictions. On the other hand, it would be enormously useful to be able to assign semantics to the R768 dimensions of the embeddings. This would considerably enrich the explanatory capabilities of PhenoLinker on its gene-phenotype associations on top of what we already do with gene attributes (see Fig. \ref{fig:4}, panel b and the Web app). We believe that exploring BioBERT's explainability could offer valuable insights and enhance the quality of our work.

For the training and evaluation of this model, a server with an NVIDIA Tesla A100 with 40GB of VRAM, an Intel(R) Xeon Silver 4316 CPU @ 2.30GHz and 78 GB of RAM has been used. On this hardware, the average time of a training epoch is 30 seconds, model inference is on the order of milliseconds and the production time of an interpretation is in the order of seconds. The size of our model is shown in the Supplementary Fig. \ref{fig:s2}, where we can see the number of parameters and the layers of the model.

\section{Conclusions}

We have created a heterogeneous structure graph neural network called PhenoLinker specifically tailored for predicting gene-phenotype associations using HPO, and it has exhibited superior performance compared to previously published models. The model's predictions, scores, explanations, and various visualization methods are accessible to the research community via a Hugging Face Space (link). Finally, our model's gene-phenotype predictions have contributed to enhanced genetic diagnoses in real-life cases.

\bibliographystyle{ieeetr}
\bibliography{references}  






\pagebreak
\begin{center}
\textbf{\large Supplementary Materials}
\end{center}
\setcounter{equation}{0}
\setcounter{figure}{0}
\setcounter{table}{0}
\makeatletter
\renewcommand{\theequation}{S\arabic{equation}}
\renewcommand{\thefigure}{S\arabic{figure}}
\renewcommand{\thetable}{S\arabic{table}}


\begin{table}[H]
\centering
\resizebox{\textwidth}{!}{%
\begin{tabular}{@{}
>{\columncolor[HTML]{FFFFFF}}l 
>{\columncolor[HTML]{FFFFFF}}l 
>{\columncolor[HTML]{FFFFFF}}l 
>{\columncolor[HTML]{FFFFFF}}l @{}}
\toprule
\textbf{Group} &
  \textbf{Source} &
  \textbf{Attributes} &
  \textbf{Description} \\ \midrule
\cellcolor[HTML]{FFFFFF} &
  \cellcolor[HTML]{FFFFFF} &
  ExACpLI &
  The probability of being loss-of-function, lof, intolerant of both heterozygous and homozygous lof variants \\
\cellcolor[HTML]{FFFFFF} &
  \cellcolor[HTML]{FFFFFF} &
  ExACpRec &
  The probability of being intolerant of homozygous, but not heterozygous lof variants \\
\cellcolor[HTML]{FFFFFF} &
  \cellcolor[HTML]{FFFFFF} &
  ExACpNull &
  The probability of being tolerant of both heterozygous and homozygous lof variants \\
\multirow{-4}{*}{\cellcolor[HTML]{FFFFFF}ExAC} &
  \multirow{-4}{*}{\cellcolor[HTML]{FFFFFF}The Exome Aggregation Consortium database} &
  ExACpMiss &
  Corrected missense Z score, taking into account that higher Z scores indicate that the transcript is more intolerant of variation \\ \midrule
\cellcolor[HTML]{FFFFFF} &
  \cellcolor[HTML]{FFFFFF} &
  CountsOverlap &
  The number of genes overlapping a gene of interest \\
\multirow{-2}{*}{\cellcolor[HTML]{FFFFFF}Locus} &
  \multirow{-2}{*}{\cellcolor[HTML]{FFFFFF}R package GRanges} &
  CountsProtCodOverlap &
  The number of overlapping protein coding genes \\ \midrule
Protein &
  STRING database &
  StringCombined &
  The number of catalogued interactions for the corresponding proteins \\ \midrule
\cellcolor[HTML]{FFFFFF} &
  \cellcolor[HTML]{FFFFFF} &
  constitutiveexons &
  The constituve exons of each gene \\
\cellcolor[HTML]{FFFFFF} &
  \cellcolor[HTML]{FFFFFF} &
  ESTcount &
  Number of detected ESTs \\
\cellcolor[HTML]{FFFFFF} &
  \cellcolor[HTML]{FFFFFF} &
  alt3EST &
  3' Alternative EST \\
\cellcolor[HTML]{FFFFFF} &
  \cellcolor[HTML]{FFFFFF} &
  alt5EST &
  5' Alternative EST \\
\cellcolor[HTML]{FFFFFF} &
  \cellcolor[HTML]{FFFFFF} &
  alt3.5EST &
  Alternative 3' and 5' EST \\
\cellcolor[HTML]{FFFFFF} &
  \cellcolor[HTML]{FFFFFF} &
  GeneLength &
  Gene length \\
\cellcolor[HTML]{FFFFFF} &
  \cellcolor[HTML]{FFFFFF} &
  TranscriptCount &
  Number of transcripts for the corresponding gene \\
\cellcolor[HTML]{FFFFFF} &
  \cellcolor[HTML]{FFFFFF} &
  GCcontent &
  Percentage of GC content \\
\cellcolor[HTML]{FFFFFF} &
  \cellcolor[HTML]{FFFFFF} &
  NumJunctions &
  The total number of unique exon-exon junctions generated by each gene \\
\multirow{-10}{*}{\cellcolor[HTML]{FFFFFF}Genomic} &
  \multirow{-10}{*}{\cellcolor[HTML]{FFFFFF}\begin{tabular}[c]{@{}l@{}}"Ensembl\\ version 72"\end{tabular}} &
  IntronicLength &
  The intronic length of each gene \\ \midrule
Expression &
  GTEx V6 gene expression dataset &
  47 tissue expression attributes &
  Boolean that indicates if the gene is expressed in the tissue or not \\ \bottomrule
\end{tabular}%
}
\vspace{1mm}
\caption{Attribute groups for genes, where we can see the sources for each group, the attributes and a description for each one.}
\label{tab:s1}
\end{table}

\begin{table}[H]
\centering
\begin{tabular}{@{}ll@{}}
\toprule
\textbf{Phenotype description}       & \textbf{HPO ID} \\ \midrule
Cognitive impairment                 & HP:0100543      \\
Mental deterioration                 & HP:0001268      \\
Intellectual disability              & HP:0001249      \\
Global developmental delay           & HP:0001263      \\
Intellectual disability, severe      & HP:0010864      \\
Intellectual disability, progressive & HP:0006887      \\
Intellectual disability, profound    & HP:0002187      \\
Intellectual disability, moderate    & HP:0002342      \\
Intellectual disability, mild        & HP:0001256      \\
Intellectual disability, borderline  & HP:0006889      \\ \bottomrule
\end{tabular}%
\vspace{1mm}
\caption{List of HPO phenotype terms associated with intellectual disability.}
\label{tab:s2}
\end{table}

\begin{table}[H]
\centering
\resizebox{\textwidth}{!}{%
\begin{tabular}{@{}
>{\columncolor[HTML]{FFFFFF}}l 
>{\columncolor[HTML]{FFFFFF}}l 
>{\columncolor[HTML]{FFFFFF}}l 
>{\columncolor[HTML]{FFFFFF}}l 
>{\columncolor[HTML]{FFFFFF}}l 
>{\columncolor[HTML]{FFFFFF}}l @{}}
\toprule
\textbf{Disease Panel} &
  \textbf{Relations} &
  \textbf{Best Threshold Disease Related} &
  \textbf{Best Fold Enrichment Disease Related} &
  \textbf{Best Threshold Full Randomized} &
  \textbf{Best Fold Enrichment Full Randomized} \\ \midrule
Developmental Disorder (DD) & 3030 & 0,9  & 4,9 & 0,85 & 14   \\
Skeletal                    & 436  & 0,95 & 2,5 & 0,85 & 19   \\
Eye                         & 1470 & 0,93 & 3,7 & 0,93 & 15,1 \\
Skin                        & 767  & 0,92 & 3   & 0,9  & 16,3 \\ \bottomrule
\end{tabular}%
}
\vspace{1mm}
\caption{Fold enrichment analysis. For each panel, the maximum fold reached and the threshold where it is reached are shown. This analysis is performed both by comparing with disease-related random distributions and with completely random distributions.}
\label{tab:s3}
\end{table}

\begin{figure}[H]
    \centering
    \includegraphics[width=\textwidth]{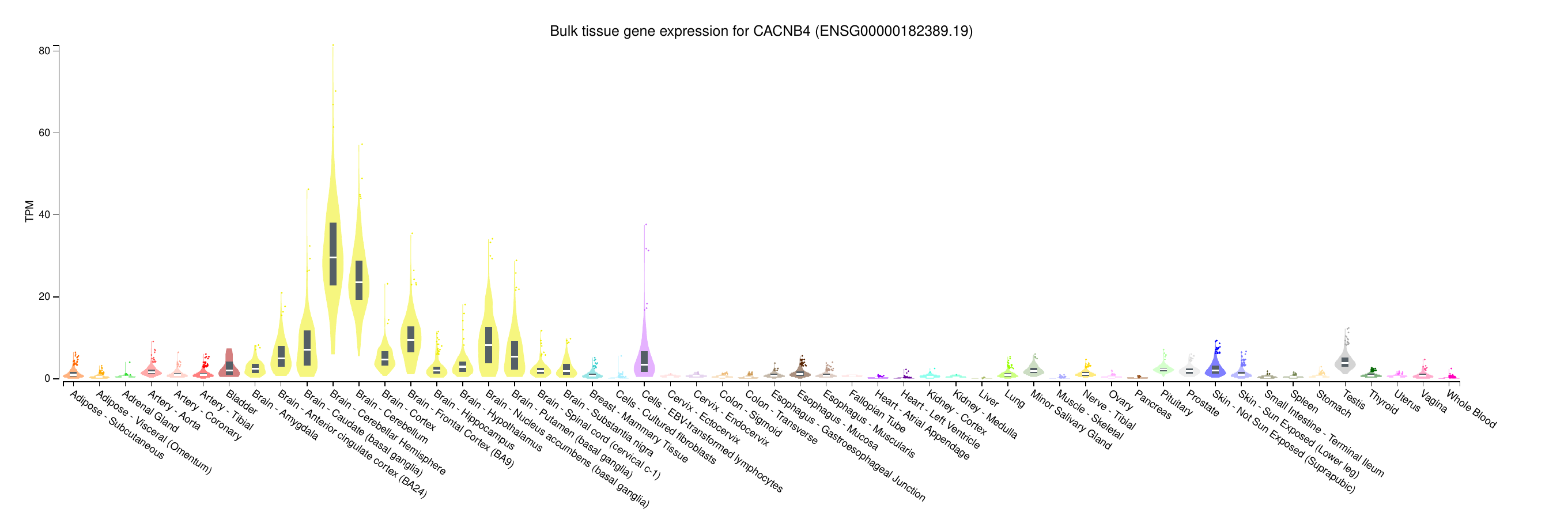}
    \caption{Bulk tissue gene expression for gene CACNB4, where it is shown that the gene is mainly expressed in the cerebellar hemisphere.}
    \label{fig:s1}
\end{figure}

\begin{figure}[H]
    \centering
    \includegraphics[width=\textwidth]{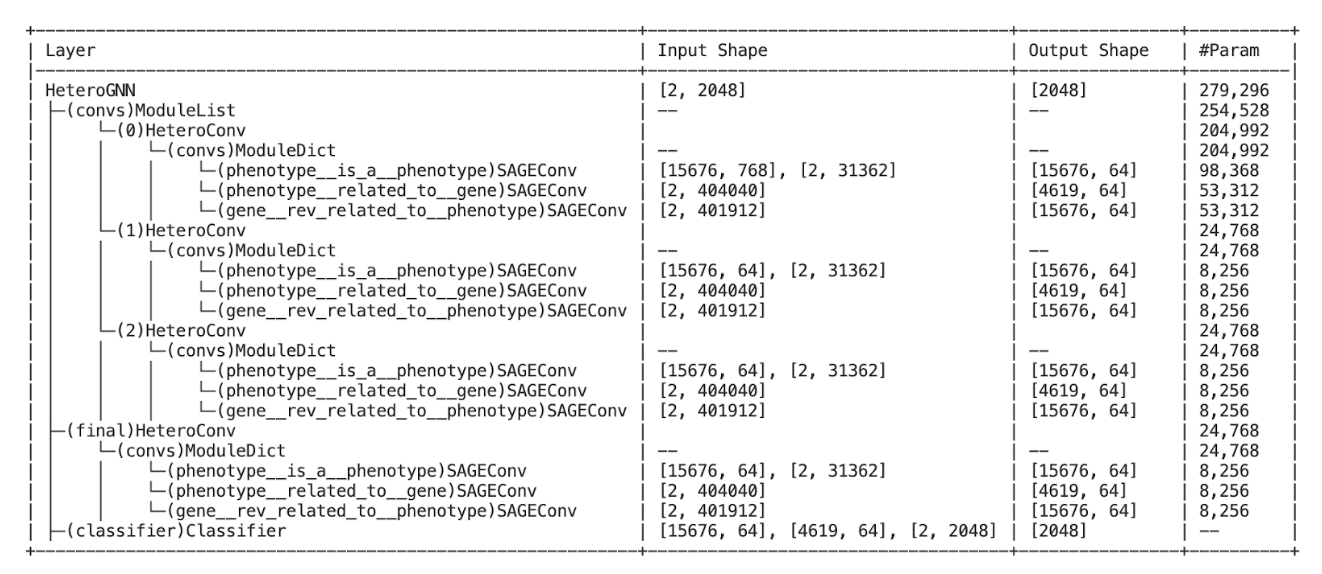}
    \caption{Summary of the PhenoLinker model. The layers, input/output shapes and number of parameters are shown for an inference on the test set with a batch of 2048 links.}
    \label{fig:s2}
\end{figure}

\begin{longtable}[c]{@{}
>{\columncolor[HTML]{FFFFFF}}l 
>{\columncolor[HTML]{FFFFFF}}l 
>{\columncolor[HTML]{FFFFFF}}l 
>{\columncolor[HTML]{FFFFFF}}l @{}}
\toprule
{\color[HTML]{000000} \textbf{Gene}} & {\color[HTML]{000000} \textbf{Phenotype}} & {\color[HTML]{000000} \textbf{Name}} & {\color[HTML]{000000} \textbf{Score}} \\* \midrule
\endfirsthead
\endhead
\bottomrule
\endfoot
\endlastfoot
ACADSB   & HP:0001249 & Intellectual disability           & 0,9821 \\
ACADVL   & HP:0001249 & Intellectual disability           & 0,9540 \\
ACADVL   & HP:0001263 & Global developmental delay        & 0,9672 \\
ACAT1    & HP:0100543 & Cognitive impairment              & 0,9037 \\
ACTA2    & HP:0001263 & Global developmental delay        & 0,9839 \\
ACVR1    & HP:0001263 & Global developmental delay        & 0,9285 \\
ADAMTS10 & HP:0001263 & Global developmental delay        & 0,9875 \\
AIMP2    & HP:0010864 & Intellectual disability, severe   & 0,9079 \\
AIMP2    & HP:0100543 & Cognitive impairment              & 0,9314 \\
ALDOA    & HP:0001263 & Global developmental delay        & 0,9156 \\
ALPL     & HP:0001249 & Intellectual disability           & 0,9369 \\
ALPL     & HP:0001263 & Global developmental delay        & 0,9626 \\
ALX3     & HP:0001263 & Global developmental delay        & 0,9944 \\
ANTXR1   & HP:0001263 & Global developmental delay        & 0,9864 \\
ATP11A   & HP:0001249 & Intellectual disability           & 0,9785 \\
B4GALT1  & HP:0001249 & Intellectual disability           & 0,9902 \\
BSND     & HP:0001263 & Global developmental delay        & 0,9942 \\
CACNA2D2 & HP:0001249 & Intellectual disability           & 0,9930 \\
CACNA2D2 & HP:0001268 & Mental deterioration              & 0,9043 \\
CACNA2D2 & HP:0100543 & Cognitive impairment              & 0,9125 \\
CACNB4   & HP:0001249 & Intellectual disability           & 0,9306 \\
CACNB4   & HP:0001263 & Global developmental delay        & 0,9497 \\
CACNB4   & HP:0100543 & Cognitive impairment              & 0,9673 \\
CASR     & HP:0001249 & Intellectual disability           & 0,9455 \\
CASR     & HP:0001263 & Global developmental delay        & 0,9408 \\
CCDC88A  & HP:0010864 & Intellectual disability, severe   & 0,9540 \\
CCDC88A  & HP:0100543 & Cognitive impairment              & 0,9559 \\
CDKN1C   & HP:0001249 & Intellectual disability           & 0,9861 \\
CLCN2    & HP:0001249 & Intellectual disability           & 0,9090 \\
CLCN2    & HP:0001263 & Global developmental delay        & 0,9191 \\
CNPY3    & HP:0100543 & Cognitive impairment              & 0,9339 \\
COPB1    & HP:0001263 & Global developmental delay        & 0,9943 \\
COPB1    & HP:0100543 & Cognitive impairment              & 0,9015 \\
COQ9     & HP:0001249 & Intellectual disability           & 0,9870 \\
CPSF3    & HP:0010864 & Intellectual disability, severe   & 0,9237 \\
CPSF3    & HP:0100543 & Cognitive impairment              & 0,9328 \\
CTC1     & HP:0001249 & Intellectual disability           & 0,9734 \\
CYP27A1  & HP:0001263 & Global developmental delay        & 0,9893 \\
CYP2U1   & HP:0001268 & Mental deterioration              & 0,9022 \\
DALRD3   & HP:0001256 & Intellectual disability, mild     & 0,9088 \\
DALRD3   & HP:0010864 & Intellectual disability, severe   & 0,9366 \\
DDOST    & HP:0001249 & Intellectual disability           & 0,9617 \\
DENND5A  & HP:0001249 & Intellectual disability           & 0,9921 \\
DENND5A  & HP:0010864 & Intellectual disability, severe   & 0,9354 \\
DENND5A  & HP:0100543 & Cognitive impairment              & 0,9431 \\
DLG1     & HP:0001263 & Global developmental delay        & 0,9879 \\
DOHH     & HP:0001268 & Mental deterioration              & 0,9284 \\
DPH2     & HP:0001249 & Intellectual disability           & 0,9820 \\
DPM3     & HP:0001263 & Global developmental delay        & 0,9366 \\
DYNC1I2  & HP:0100543 & Cognitive impairment              & 0,9342 \\
EMG1     & HP:0001249 & Intellectual disability           & 0,9681 \\
ENTPD1   & HP:0001263 & Global developmental delay        & 0,9935 \\
EXOC2    & HP:0001249 & Intellectual disability           & 0,9951 \\
EXOC2    & HP:0010864 & Intellectual disability, severe   & 0,9307 \\
EXOC7    & HP:0001249 & Intellectual disability           & 0,9834 \\
EXOC7    & HP:0010864 & Intellectual disability, severe   & 0,9051 \\
EXOSC8   & HP:0001249 & Intellectual disability           & 0,9948 \\
EXOSC8   & HP:0001268 & Mental deterioration              & 0,9011 \\
EXOSC8   & HP:0100543 & Cognitive impairment              & 0,9544 \\
FARSB    & HP:0001249 & Intellectual disability           & 0,9761 \\
FDFT1    & HP:0001249 & Intellectual disability           & 0,9879 \\
FGF13    & HP:0010864 & Intellectual disability, severe   & 0,9321 \\
FGF14    & HP:0001263 & Global developmental delay        & 0,9622 \\
FGF14    & HP:0001268 & Mental deterioration              & 0,9503 \\
FRAS1    & HP:0001263 & Global developmental delay        & 0,9954 \\
FREM2    & HP:0001263 & Global developmental delay        & 0,9929 \\
FRMD5    & HP:0001268 & Mental deterioration              & 0,9123 \\
FRMD5    & HP:0100543 & Cognitive impairment              & 0,9756 \\
FTO      & HP:0001249 & Intellectual disability           & 0,9910 \\
GBA2     & HP:0001263 & Global developmental delay        & 0,9940 \\
GCSH     & HP:0001263 & Global developmental delay        & 0,9815 \\
GCSH     & HP:0100543 & Cognitive impairment              & 0,9410 \\
GEMIN4   & HP:0001249 & Intellectual disability           & 0,9805 \\
GJB1     & HP:0001249 & Intellectual disability           & 0,9662 \\
GJB1     & HP:0001263 & Global developmental delay        & 0,9620 \\
GJB3     & HP:0001263 & Global developmental delay        & 0,9622 \\
GLS      & HP:0001249 & Intellectual disability           & 0,9912 \\
GLS      & HP:0001268 & Mental deterioration              & 0,9296 \\
GLS      & HP:0100543 & Cognitive impairment              & 0,9644 \\
GOT2     & HP:0001263 & Global developmental delay        & 0,9952 \\
GPSM2    & HP:0001263 & Global developmental delay        & 0,9795 \\
GRIA1    & HP:0100543 & Cognitive impairment              & 0,9540 \\
GRM7     & HP:0001256 & Intellectual disability, mild     & 0,9193 \\
GRM7     & HP:0002342 & Intellectual disability, moderate & 0,9058 \\
GRM7     & HP:0010864 & Intellectual disability, severe   & 0,9439 \\
GRM7     & HP:0100543 & Cognitive impairment              & 0,9739 \\
GSS      & HP:0001263 & Global developmental delay        & 0,9312 \\
GSX2     & HP:0001249 & Intellectual disability           & 0,9912 \\
GSX2     & HP:0001268 & Mental deterioration              & 0,9305 \\
GSX2     & HP:0100543 & Cognitive impairment              & 0,9125 \\
HADHB    & HP:0001249 & Intellectual disability           & 0,9670 \\
HEATR3   & HP:0001263 & Global developmental delay        & 0,9729 \\
IFT27    & HP:0001263 & Global developmental delay        & 0,9359 \\
IFT43    & HP:0001249 & Intellectual disability           & 0,9597 \\
IFT43    & HP:0001263 & Global developmental delay        & 0,9659 \\
IREB2    & HP:0001249 & Intellectual disability           & 0,9896 \\
IREB2    & HP:0100543 & Cognitive impairment              & 0,9469 \\
ISCA2    & HP:0001249 & Intellectual disability           & 0,9904 \\
ISCA2    & HP:0001263 & Global developmental delay        & 0,9962 \\
ISCA2    & HP:0100543 & Cognitive impairment              & 0,0946 \\
ITGA7    & HP:0001263 & Global developmental delay        & 0,9977 \\
KCNC3    & HP:0001268 & Mental deterioration              & 0,9514 \\
KIF4A    & HP:0001263 & Global developmental delay        & 0,9729 \\
LARS2    & HP:0001249 & Intellectual disability           & 0,9686 \\
LARS2    & HP:0001263 & Global developmental delay        & 0,9817 \\
LIPT2    & HP:0001249 & Intellectual disability           & 0,9920 \\
LIPT2    & HP:0001268 & Mental deterioration              & 0,9391 \\
LIPT2    & HP:0100543 & Cognitive impairment              & 0,9033 \\
LNPK     & HP:0001249 & Intellectual disability           & 0,0985 \\
MAN2C1   & HP:0001263 & Global developmental delay        & 0,9922 \\
MAPKAPK5 & HP:0001249 & Intellectual disability           & 0,9872 \\
MED12L   & HP:0001263 & Global developmental delay        & 0,9983 \\
MPV17    & HP:0001249 & Intellectual disability           & 0,9597 \\
NAGS     & HP:0001249 & Intellectual disability           & 0,9397 \\
NBAS     & HP:0001249 & Intellectual disability           & 0,9720 \\
NBAS     & HP:0001263 & Global developmental delay        & 0,9835 \\
NBN      & HP:0001263 & Global developmental delay        & 0,9857 \\
NCAPD2   & HP:0001263 & Global developmental delay        & 0,0984 \\
NCAPG2   & HP:0001249 & Intellectual disability           & 0,9921 \\
NDUFAF1  & HP:0001249 & Intellectual disability           & 0,9929 \\
NDUFAF1  & HP:0100543 & Cognitive impairment              & 0,9037 \\
NDUFAF2  & HP:0001268 & Mental deterioration              & 0,9626 \\
NDUFAF2  & HP:0100543 & Cognitive impairment              & 0,0982 \\
NDUFAF5  & HP:0001268 & Mental deterioration              & 0,0962 \\
NDUFAF5  & HP:0100543 & Cognitive impairment              & 0,9754 \\
NECAP1   & HP:0001256 & Intellectual disability, mild     & 0,9380 \\
NECAP1   & HP:0010864 & Intellectual disability, severe   & 0,0931 \\
NFIB     & HP:0001249 & Intellectual disability           & 0,9537 \\
NFIB     & HP:0001263 & Global developmental delay        & 0,9796 \\
NPHP3    & HP:0001249 & Intellectual disability           & 0,9645 \\
NUP188   & HP:0001249 & Intellectual disability           & 0,9862 \\
NUP188   & HP:0001263 & Global developmental delay        & 0,9964 \\
NUP214   & HP:0001249 & Intellectual disability           & 0,0959 \\
NUP62    & HP:0001263 & Global developmental delay        & 0,9959 \\
NUP62    & HP:0001268 & Mental deterioration              & 0,9730 \\
NUP62    & HP:0100543 & Cognitive impairment              & 0,9743 \\
PAM16    & HP:0001249 & Intellectual disability           & 0,9423 \\
PDE10A   & HP:0001263 & Global developmental delay        & 0,9713 \\
PIK3C2A  & HP:0001249 & Intellectual disability           & 0,9808 \\
PLXNA1   & HP:0001249 & Intellectual disability           & 0,9885 \\
PNPO     & HP:0001249 & Intellectual disability           & 0,9914 \\
POMK     & HP:0010864 & Intellectual disability, severe   & 0,9448 \\
PPFIBP1  & HP:0010864 & Intellectual disability, severe   & 0,9259 \\
PRKACB   & HP:0001263 & Global developmental delay        & 0,9870 \\
PRKAR1B  & HP:0001249 & Intellectual disability           & 0,9935 \\
PRKAR1B  & HP:0001256 & Intellectual disability, mild     & 0,9154 \\
PRKD1    & HP:0001249 & Intellectual disability           & 0,9791 \\
PRRT2    & HP:0001263 & Global developmental delay        & 0,9863 \\
PRRT2    & HP:0001268 & Mental deterioration              & 0,9842 \\
PRRT2    & HP:0100543 & Cognitive impairment              & 0,9937 \\
PSMB8    & HP:0001263 & Global developmental delay        & 0,9835 \\
PTH1R    & HP:0001249 & Intellectual disability           & 0,9678 \\
PTH1R    & HP:0001263 & Global developmental delay        & 0,9826 \\
RBPJ     & HP:0001263 & Global developmental delay        & 0,9931 \\
RNF220   & HP:0001263 & Global developmental delay        & 0,9923 \\
RNF220   & HP:0001268 & Mental deterioration              & 0,9214 \\
RNF220   & HP:0100543 & Cognitive impairment              & 0,9016 \\
RPS23    & HP:0001249 & Intellectual disability           & 0,9773 \\
RPS23    & HP:0001263 & Global developmental delay        & 0,9942 \\
RSPRY1   & HP:0001263 & Global developmental delay        & 0,9886 \\
RUSC2    & HP:0001256 & Intellectual disability, mild     & 0,9198 \\
SACS     & HP:0001263 & Global developmental delay        & 0,9944 \\
SACS     & HP:0001268 & Mental deterioration              & 0,9445 \\
SACS     & HP:0100543 & Cognitive impairment              & 0,9435 \\
SCN1B    & HP:0001268 & Mental deterioration              & 0,0092 \\
SCN1B    & HP:0010864 & Intellectual disability, severe   & 0,9118 \\
SEC31A   & HP:0001249 & Intellectual disability           & 0,9899 \\
SEC31A   & HP:0001263 & Global developmental delay        & 0,9978 \\
SEMA6B   & HP:0001268 & Mental deterioration              & 0,9262 \\
SEMA6B   & HP:0100543 & Cognitive impairment              & 0,9407 \\
SHROOM4  & HP:0001263 & Global developmental delay        & 0,9927 \\
SLC26A2  & HP:0001249 & Intellectual disability           & 0,9675 \\
SLC26A2  & HP:0001263 & Global developmental delay        & 0,9840 \\
SLC35D1  & HP:0001263 & Global developmental delay        & 0,9137 \\
SLC39A13 & HP:0001249 & Intellectual disability           & 0,9155 \\
SLC39A13 & HP:0001263 & Global developmental delay        & 0,9344 \\
SLC5A5   & HP:0001263 & Global developmental delay        & 0,9794 \\
SLC5A7   & HP:0001263 & Global developmental delay        & 0,9995 \\
SLC9A7   & HP:0001249 & Intellectual disability           & 0,9569 \\
SMAD3    & HP:0001249 & Intellectual disability           & 0,9838 \\
SMAD3    & HP:0001263 & Global developmental delay        & 0,9969 \\
SMG9     & HP:0010864 & Intellectual disability, severe   & 0,9150 \\
SUCLA2   & HP:0100543 & Cognitive impairment              & 0,9392 \\
TAB2     & HP:0001263 & Global developmental delay        & 0,9975 \\
TBX1     & HP:0100543 & Cognitive impairment              & 0,9128 \\
TGFB1    & HP:0001249 & Intellectual disability           & 0,9711 \\
THRB     & HP:0001249 & Intellectual disability           & 0,9328 \\
THRB     & HP:0001263 & Global developmental delay        & 0,9265 \\
TKFC     & HP:0001249 & Intellectual disability           & 0,9879 \\
TMEM147  & HP:0001263 & Global developmental delay        & 0,9985 \\
TNIK     & HP:0001263 & Global developmental delay        & 0,9879 \\
TNIK     & HP:0100543 & Cognitive impairment              & 0,9497 \\
TNR      & HP:0001249 & Intellectual disability           & 0,9923 \\
TNR      & HP:0100543 & Cognitive impairment              & 0,0915 \\
TRAPPC10                            & HP:0002342                               & Intellectual disability, moderate      & 0,9173                                       \\
TRAPPC11 & HP:0100543 & Cognitive impairment              & 0,9324 \\
TRAPPC2L & HP:0001249 & Intellectual disability           & 0,9785 \\
TSEN15   & HP:0001268 & Mental deterioration              & 0,9669 \\
TSEN15                              & HP:0002187                               & Intellectual disability, profound      & 0,9638                                       \\
TSEN15                              & HP:0006887                               & Intellectual disability, progressive   & 0,9357                                       \\
TSEN15   & HP:0010864 & Intellectual disability, severe   & 0,9618 \\
TSEN15   & HP:0100543 & Cognitive impairment              & 0,9809 \\
TSPOAP1  & HP:0001263 & Global developmental delay        & 0,9527 \\
TSPOAP1  & HP:0001268 & Mental deterioration              & 0,9634 \\
TSPOAP1  & HP:0100543 & Cognitive impairment              & 0,9796 \\
TUBGCP2  & HP:0001249 & Intellectual disability           & 0,9929 \\
TUBGCP2  & HP:0001263 & Global developmental delay        & 0,9986 \\
TUBGCP2  & HP:0010864 & Intellectual disability, severe   & 0,9300 \\
TUBGCP2  & HP:0100543 & Cognitive impairment              & 0,9392 \\
UFC1     & HP:0001263 & Global developmental delay        & 0,9983 \\
VIPAS39  & HP:0001249 & Intellectual disability           & 0,9661 \\
VPS33B   & HP:0001249 & Intellectual disability           & 0,9715 \\
VPS50    & HP:0001249 & Intellectual disability           & 0,9929 \\
VPS51    & HP:0001249 & Intellectual disability           & 0,9937 \\
VPS51    & HP:0010864 & Intellectual disability, severe   & 0,9133 \\
WASHC5   & HP:0100543 & Cognitive impairment              & 0,9016 \\
XPA      & HP:0001263 & Global developmental delay        & 0,9820 \\* \bottomrule
\caption{Gene-phenotype associations, where genes are in the intellectual disability Genomics England amber panel and phenotypes are selected from Supplementary Table \ref{tab:s1}. Only associations with a score \textgreater 0.9 are shown.}
\label{tab:s4}\\
\end{longtable}

\end{document}